# Locating the Sources of Sub-synchronous Oscillations Induced by the Control of Voltage Source Converters Based on Energy Structure and Nonlinearity Detection

Zetian Zheng, *Student Member, IEEE,* Shaowei Huang, *Member, IEEE,* Jun Yan,
Qiangsheng Bu, Chen Shen, *Senior Member, IEEE*, Mingzhong Zheng
and Ye Liu, *Student Member, IEEE*

*Abstract*—The oscillation phenomena associated with the control of voltage source converters (VSCs) are widely concerning, and locating the source of these oscillations is crucial to suppressing them; therefore, this paper presents a locating scheme, based on the energy structure and nonlinearity detection. On the one hand, the energy structure, which conforms with the principle of the energy-based method and dissipativity theory, is constructed to describe the transient energy flow for VSCs, and on this basis, a defined characteristic quantity is implemented to narrow the scope of oscillation source location; on the other hand, according to the self-sustained oscillation characteristics of VSCs, an index for nonlinearity detection is applied to locate the VSCs which produce the oscillation energy. The combination of the energy structure and nonlinearity detection could distinguish the contributions of different VSCs to the oscillation. The results of a case study implemented by the PSCAD/EMTDC simulation validate the proposed scheme.

*Index Terms*—Double-loop PI control, energy structure, Hamiltonian model, nonlinearity detection, oscillation source location (OSL), Voltage Source Converter (VSC).

## Nomenclature

| | |
|---|---|
| $H_{VL}$, $H_{CL}$ | Hamiltonian storage functions for the voltage and current loop control of the VSC, respectively. |
| $E_{SVG1}$, $E_{SVG2}$ | Output potentials for $SVG_1$ and $SVG_2$, respectively. |
| $THD_U$, $THD_I$ | Total harmonic distortion (THD) for voltage and current, respectively. |

## I. Introduction

IN modern power systems, voltage source converters (VSCs) are among the most common power electronic devices. Typical application scenarios of VSCs range from the renewable energy generation, such as wind farms, to high-voltage DC and flexible AC transmission systems [1]-[3]. The existing literature [3]–[8] shows that with the high penetration of power converters, the dynamic characteristics of the power system have undergone significant changes, so that new oscillatory phenomena have emerged, threatening system stability. Some of these phenomena are associated with VSC control.

Across industry and academia, there is a consensus that locating the sources of oscillation is an important measure to suppress oscillation [9], [10]. Correspondingly, in [11], numerous methods for oscillation source location (OSL) were surveyed and categorized; the most notable is the energy-based method (EBM) [12], which tracks the system-wide energy flow to locate the oscillation sources. The advantages of the EBM include the following: (i) compared to the location methods based on damping torque analysis or mode shape estimation, the EBM is adapted to locate forced oscillations as well as poorly damped oscillations [11]; and (ii) the EBM is convenient for voltage/current measurements in wide-area networks [12], [13]. With the rapid development of phasor measurement units (PMU), the EBM has been successfully used for oscillation monitoring in actual power systems [14]. Thus, this study focuses on the EBM considering its prospects for industrial applications.

In recent years, the EBM has been developed. For example, Wu *et al.* [13] proposed a distributed cooperative scheme to locate a forced oscillation source by detecting a cut-set energy flow. In addition, some studies focused on the oscillations associated with wind farms. Ma *et al.* [15] developed an equipment-level locating method for low-frequency oscillation sources in power systems with doubly fed induction generator (DFIG) integration, based on an energy correlation topology network and dynamic energy flow. Lei *et al.* [16] presented a forced oscillation source location and participation assessment method for DFIGs by analyzing the energy flow, and based on this analysis, the participation factor for oscillations is proposed. Reviewing [13], [15], and [16], the formulas for energy flow, which are suitable for low-frequency OSL, were derived from [12] and proven in [17] and [18] to conform with the dissipativity theory. However, according to [4] and [6], there is a risk of sub-synchronous oscillation (SSO) in multi-VSC systems; therefore, the analysis of transient energy flow (TEF)

---

This work was supported by the State Grid Guide Project under Grant No. 5108-202218030A-1-1-ZN. (Corresponding author: Chen Shen.)

Z. Zheng, S. Huang, J. Yan, C. Shen and Y. Liu are with the State Key Laboratory of Power Systems, Department of Electrical Engineering, Tsinghua University, Beijing 100084, China (e-mail: zzt_thu@qq.com; huangsw@tsinghua.edu.cn; 458783027@qq.com; shenchen@mail.tsing-hua.edu.cn; liuye18@mails.tsinghua.edu.cn).

Q. Bu and M. Zheng are with State Grid Jiangsu Electric Power Research Institute, Nanjing, Jiangsu Province 211103, China (e-mail: tc16002315@163.com; mingzhongz@tju.edu.cn).

under the SSO condition is crucial for locating the sources of oscillations caused by VSC control.

To date, the problem of TEF has attracted the interest of several researchers. In [19], a criterion for identifying the source of SSO was proposed based on TEF, considering the control interaction between converter-based wind turbine generators and weak AC grids. The authors of [20] focusing on the SSO associated with the control of the VSC in direct-drive wind farms, presented the formulas for TEF, and studied the relationship between the energy flow and damping torque. Unfortunately, the formulas for TEF presented in the above literature are not accompanied by proof of conformance to the dissipativity theory. That is, the problem of TEF has not yet been fully addressed.

In addition, some researchers have been concerned about the role of the VSC nonlinear characteristics in the SSO. According to [4] and [21], a VSC-based wind farm may cause a sustained SSO when the control limit is satisfied; evidently, the control limit possesses nonlinear characteristics. In [21], the describing function method was adopted to analyze the nonlinearity of the VSC control limit; however, the authors did not discuss the relevant nonlinearity detection and OSL for the oscillation associated with the limitation of the VSC control. In addition, the authors of [19] classified the external characteristics of the above-mentioned oscillation as an inter-harmonic source and constructed a framework for SSO identification considering the harmonic characteristics; however, the result obtained in [19] does not support the distinction of oscillation responsibility among multiple sources.

Therefore, this paper presents a scheme for the source location of SSO caused by the control of VSCs, based on the energy structure and nonlinearity detection. On the one hand, the energy structure, which is based on the principle of the EBM and conforms with the dissipativity theory, is constructed to describe the TEF for VSCs; on the other hand, the method of nonlinearity detection is applied to aid in the identification of the oscillation responsibility. The main contributions of this study are as follows.

1) An equipment-level energy structure is established based on the Hamiltonian model for the VSC, including the main circuit and control loops. Based on the energy structure, oscillation monitoring can be implemented using a defined characteristic quantity derived from the energy function. Furthermore, this characteristic quantity can be obtained easily using instantaneous voltage/current measurements.

2) The voltage disturbance can induce the nonlinear oscillation, considering the limitation characteristic of VSC control, and an index $\mu$ for nonlinearity detection is introduced to distinguish the nonlinear oscillation from the linear one.

3) Targeting the SSO caused by VSC control, a novel scheme for OSL is proposed. First, the instantaneous voltages and currents are measured to monitor the oscillation; once the oscillation occurs, the oscillation source researching based on the characteristic quantity $\Delta ESP_{ac}$ is triggered to determine the zones or nodes that produce the oscillating energy, in order to narrow the node scope for the OSL. The Hamiltonian energy functions $H_{VL}$ and $H_{CL}$ for the VSC control are used to further identify and locate the oscillation source. Finally, nonlinearity detection is applied to determine the type of oscillation and identify the oscillation responsibility.

The remainder of this paper is organized as follows: Section II introduces the energy structure for VSC. Section III discusses the methodology of nonlinearity detection for determining the oscillation type. Section IV proposes a scheme for OSL, whereas the feasibility of the proposed scheme, discussed using a case study, is described in Section V. Finally, conclusions are drawn in Section VI.

## II. Hamiltonian Model and Energy Structure for VSC

This section presents the port-controlled Hamiltonian (PCH) model for the VSC and illustrates the corresponding energy structure, which is the basis for monitoring and locating the sources of oscillations.

### A. The Main Circuit of VSC

Fig. 1 shows the most common topology of the three-phase VSC, where, $v_a$, $v_b$, and $v_c$ are the three-phase voltages at the point of common coupling (PCC); $e_a$, $e_b$, and $e_c$ are the AC-side voltages of the VSC, and $v_{dc}$ is the DC-side voltage; $i_a$, $i_b$, and $i_c$ are the three-phase currents of the VSC, and $i_{dc}$ is the DC-side current; $R_{ac}$ and $L_{ac}$ are the AC-side equivalent resistance and reactance, respectively. The state equations for the main circuit of the VSC can be described as

$$L_{ac}\frac{di_j}{dt} = e_j - v_j - R_{ac}i_j \quad (j=a,b,c)$$
$$C_{dc}\frac{dv_{dc}}{dt} = -(s_a \cdot i_a + s_b \cdot i_b + s_c \cdot i_c) + i_{dc}. \tag{1}$$

where $s_a$, $s_b$, and $s_c$ are the switching functions of the VSC. If $s_j = 1$, the upper arm of phase $j$ is turned on and the lower arm is turned off; in contrast, if $s_j = 0$, the situation is the opposite.

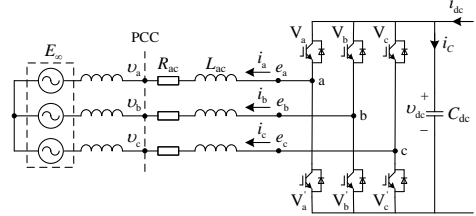

Fig. 1. Topology of three-phase VSC.

According to the principle of pulse-width modulation and Hamiltonian theory, if we have

$$e_j = s_j v_{dc} - (s_a + s_b + s_c)v_{dc}/3 = T_j v_{dc}, \tag{2}$$

then Eq.(1) could be turned into the form

$$\dot{x} = (\mathbf{J} - \mathbf{R})\frac{\partial H(x)}{\partial x} + \mathbf{G}u$$
$$y = \mathbf{G}^T \frac{\partial H(x)}{\partial x} \tag{3}$$

where

$$x = [x_1 \ x_2 \ x_3 \ x_4]^T = [L_{ac}i_a \ L_{ac}i_b \ L_{ac}i_c \ C_{dc}v_{dc}]^T \tag{4}$$

$$H(x) = x_1^2/2L_{ac} + x_2^2/2L_{ac} + x_3^2/2L_{ac} + x_4^2/2C_{dc} \tag{5}$$

$$\mathbf{J} = \begin{bmatrix} 0 & 0 & 0 & T_a \\ 0 & 0 & 0 & T_b \\ 0 & 0 & 0 & T_c \\ -T_a & -T_b & -T_c & 0 \end{bmatrix} \tag{6}$$

$$\mathbf{R} = diag[R_{ac} \ R_{ac} \ R_{ac} \ 0] \tag{7}$$

$$\mathbf{G} = diag\begin{bmatrix} -1 & -1 & -1 & 1 \end{bmatrix} \quad (8)$$

$$\boldsymbol{u} = \begin{bmatrix} v_a & v_b & v_c & i_{dc} \end{bmatrix}^T \quad (9)$$

In (7), the matrix **R** is positive semi-definite under normal conditions; therefore, the Hamiltonian energy function $H(x)$, as shown in (5), satisfies

$$\frac{dH(x)}{dt} = \boldsymbol{u}^T \boldsymbol{y} - \frac{\partial H^T(x)}{\partial x} \mathbf{R} \frac{\partial H(x)}{\partial x} \leq \boldsymbol{u}^T \boldsymbol{y} \quad (10)$$

Thus, the system of (3)-(9) is dissipative.

Furthermore, according to the law of energy conservation and principle of dissipativity, we can obtain the energy structure of the VSC, which shows the characteristics of the TEF through the main circuit of the VSC. Specifically, the derivation process of the energy structure is as follows:

Reviewing (5), we have the derivative function

$$\dot{H}(x) = L_{ac}i_a\dot{i}_a + L_{ac}i_b\dot{i}_b + L_{ac}i_c\dot{i}_c + C_{dc}v_{dc}\dot{v}_{dc}. \quad (11)$$

Substituting (11) into (3), we obtain

$$\begin{aligned} T_j i_j \dot{v}_{dc} - L_{ac} i_j \dot{i}_j - R_{ac} i_j^2 &= v_j i_j \quad (j = a, b, c) \\ (T_a i_a + T_b i_b + T_c i_c) \dot{v}_{dc} &= v_{dc} i_{dc} - C_{dc} v_{dc} \dot{v}_{dc}. \end{aligned} \quad (12)$$

In (12), if we set the currents $i_j$ and DC voltage $v_{dc}$ as the force variables and set the other terms such as $i_{dc}$, $C\dot{v}_{dc}$, $T_j i_j$, $L_{ac}i_j$, $R_{ac}i_j$, and $v_j$ as the flow variables, the energy structure for the main circuit of the VSC could be obtained, as shown in Fig. 2.

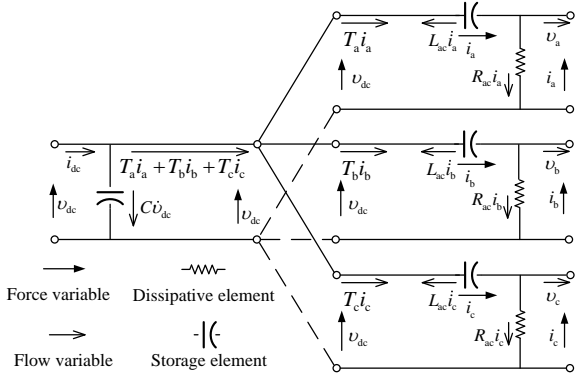

Fig. 2. Energy structure for the main circuit of the VSC.

Inspecting (11), (12), and Fig. 2, the terms $L_{ac}i_j\dot{i}_j$ and $Cv_{dc}\dot{v}_{dc}$ in the derivative function $\dot{H}(x)$ are represented by the storage elements of the energy structure; similarly, in the energy structure, the terms $-R_{ac}i_j^2$ in (12) are represented by the dissipative elements. As shown in Fig. 2, the storage energy is equal to the energy supply on port (ESP) minus the resistive dissipation energy. Therefore, according to the energy-function derivation method presented in [22], we can obtain the energy function

$$V = \frac{x_1^2}{2L_{ac}} + \frac{x_2^2}{2L_{ac}} + \frac{x_3^2}{2L_{ac}} + \frac{x_4^2}{2C_{dc}} + \int_0^\tau v_a i_a + v_b i_b + v_c i_c - v_{dc} i_{dc} dt \quad (13)$$

and

$$\begin{aligned} \dot{V} &= \frac{x_1 \dot{x}_1}{L_{ac}} + \frac{x_2 \dot{x}_2}{L_{ac}} + \frac{x_3 \dot{x}_3}{L_{ac}} + \frac{x_4 \dot{x}_4}{C_{dc}} + v_a i_a + v_b i_b + v_c i_c - v_{dc} i_{dc} \\ &= -R_{ac}(x_1^2 + x_2^2 + x_3^2)/L_{ac}^2 \leq 0. \end{aligned} \quad (14)$$

Evidently, the energy function $V$ can be used to determine the stability of the VSC system; for example, if the equivalent resistance $R_{ac}$ changes from positive to negative damping under abnormal conditions, then $\dot{V} > 0$, and the VSC produces oscillation energy. Therefore, according to (13), to investigate the influence of the VSC on AC grids, we can describe the AC-side TEF as

$$TEF_{ac} = \int_0^\tau v_a i_a + v_b i_b + v_c i_c dt = \int_0^\tau p_{ac} dt. \quad (15)$$

Actually, (15) is equivalent to the ESP at the AC port of the VSC, as shown in Fig. 2. Although the form of (15) is simple, it is derived from the PCH model, which conforms to the dissipativity principle. Furthermore, to monitor the oscillation energy at the AC-side of the VSC, we can define the characteristic quantity $\Delta ESP_{ac}$:

$$\Delta ESP_{ac} = \int_0^\tau p_{ac} - \bar{p}_{ac} dt. \quad (16)$$

where $\Delta ESP_{ac}$ represents the accumulation of oscillation energy, and it is suitable for PMU measurement.

### B. Double-PI Control Loops of VSC

In the most common applications, the VSCs generally adopt the double-loop PI control strategy composed of an outer voltage loop and an inner current loop, as shown in Fig. 3. Fig. 3(a) shows the block diagrams of the voltage controllers, and according to the block diagrams, we can obtain the corresponding PCH model expressed as follows:

$$\dot{\boldsymbol{x}} = \begin{bmatrix} \dot{x}_5 \\ \dot{x}_6 \end{bmatrix} = \begin{bmatrix} -\frac{1}{K_{Pvd}} & 0 \\ 0 & -\frac{1}{K_{Pvq}} \end{bmatrix} \begin{bmatrix} \frac{\partial H_{VL}}{\partial x_5} \\ \frac{\partial H_{VL}}{\partial x_6} \end{bmatrix} + \begin{bmatrix} \frac{1}{K_{Pvd}} & 0 \\ 0 & \frac{1}{K_{Pvq}} \end{bmatrix} \boldsymbol{u} \quad (17)$$

$$\boldsymbol{y} = \begin{bmatrix} \frac{1}{K_{Pvd}} \frac{\partial H_{VL}}{\partial x_5} & \frac{1}{K_{Pvq}} \frac{\partial H_{VL}}{\partial x_6} \end{bmatrix}^T$$

where

$$x_5 = \int (v_{dcref} - v_{dc}) dt \quad x_6 = \int (v_{dref} - v_d) dt \quad \boldsymbol{u} = \begin{bmatrix} i_{dref} & i_{qref} \end{bmatrix}^T \quad (18)$$

$$H_{VL}(x) = K_{Ivd} x_5^2 / 2 + K_{Ivq} x_6^2 / 2. \quad (19)$$

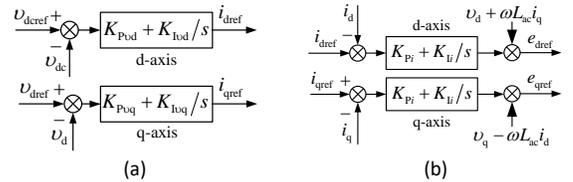

(a)   (b)

Fig. 3. Block diagrams for voltage (a) and current (b) control loops, where $v_{dcref}$ and $v_{dref}$ are the DC and d-axis reference voltages, respectively; $i_{dref}$ and $i_{qref}$ are the d-axis and q-axis reference currents, respectively; $K_{Pvd}$, $K_{Pvq}$, and $K_{Pi}$ are the proportion gains; and $K_{Ivd}$, $K_{Ivq}$, and $K_{Ii}$ are the integral gains.

From (17)–(19), we obtain

$$\frac{dH_{VL}(x)}{dt} = \boldsymbol{u}^T \boldsymbol{y} - \begin{bmatrix} \frac{\partial H_{VL}}{\partial x_5} \\ \frac{\partial H_{VL}}{\partial x_6} \end{bmatrix}^T \begin{bmatrix} -\frac{1}{K_{Pvd}} & 0 \\ 0 & -\frac{1}{K_{Pvq}} \end{bmatrix} \begin{bmatrix} \frac{\partial H_{VL}}{\partial x_5} \\ \frac{\partial H_{VL}}{\partial x_6} \end{bmatrix} \leq \boldsymbol{u}^T \boldsymbol{y}. \quad (20)$$

Hence, the voltage-loop control of the VSC is dissipative. Similarly to Section II-A, the corresponding energy structure for the voltage control can be obtained using the PCH model (17), as shown in Fig. 4(a). Similarly, according to the block diagrams of the current-loop control shown in Fig. 3(b), if

$$x_7 = \int (i_{dref} - i_d) dt \quad x_8 = \int (i_{qref} - i_q) dt \quad (21)$$

$$H_{CL}(\boldsymbol{x}) = K_{Ii} x_7^2/2 + K_{Ii} x_8^2/2 \quad (22)$$

the corresponding PCH model could be described as

$$\begin{bmatrix} \dot{x}_7 \\ \dot{x}_8 \end{bmatrix} = (\mathbf{J} - \mathbf{R}) \begin{bmatrix} \dfrac{\partial H_{CL}}{\partial x_7} \\ \dfrac{\partial H_{CL}}{\partial x_8} \end{bmatrix} + \begin{bmatrix} c & d \\ -d & c \end{bmatrix} \begin{bmatrix} \sigma_d \\ \sigma_q \end{bmatrix} \quad (23)$$

$$\boldsymbol{y} = \begin{bmatrix} c\dfrac{\partial H_{CL}}{\partial x_7} - d\dfrac{\partial H_{CL}}{\partial x_8} & d\dfrac{\partial H_{CL}}{\partial x_7} + c\dfrac{\partial H_{CL}}{\partial x_8} \end{bmatrix}^T \quad (24)$$

where

$$\mathbf{J} = \begin{bmatrix} 0 & -b/K_{Ii} \\ b/K_{Ii} & 0 \end{bmatrix} \quad \mathbf{R} = \begin{bmatrix} a/K_{Ii} & 0 \\ 0 & a/K_{Ii} \end{bmatrix} \quad (25)$$

$$\sigma_d = e_{dref} - v_d + \omega L_{ac} i_{qref} \quad \sigma_q = e_{qref} - v_q - \omega L_{ac} i_{dref} \quad (26)$$

$$a = \dfrac{K_{Ii} K_{Pi}}{\omega^2 L_{ac}^2 + K_{Pi}^2} \quad b = \dfrac{\omega L_{ac} \cdot K_{Ii}}{\omega^2 L_{ac}^2 + K_{Pi}^2} \quad c = \dfrac{a}{K_{Ii}} \quad d = \dfrac{b}{K_{Ii}}. \quad (27)$$

From (23)–(27), we can also obtain the energy structure shown in Fig. 4(b) and prove the dissipativity of the current-loop control.

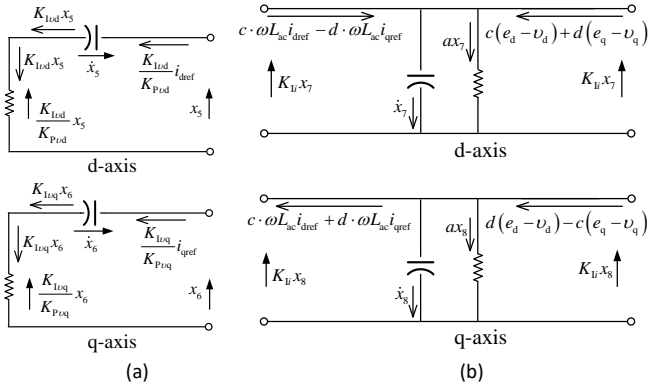

Fig. 4. Energy structure for the voltage (a) and current (b) control of the VSC. The energy structure is consistent with the control strategy, which includes two parts: d-axis control and q-axis control.

In view of the dissipativity theory, the Hamiltonian storage functions $H_{VL}$ and $H_{CL}$ conform to the essentials of the Lyapunov function [23]. In other words, the continuous increase in $H_{VL}$ and $H_{CL}$ could represent an unstable accumulation of energy accompanied by oscillation. Thus, $H_{VL}$ and $H_{CL}$ can be regarded as characteristic quantities for identifying the oscillation caused by the VSC control.

### III. METHODOLOGY FOR NONLINEARITY DETECTION

Section II-A presented the characteristic quantity (16) used to monitor the oscillation energy and locate the SSO sources. However, in practice, it is necessary not only to locate the oscillation sources but also to identify the linear or nonlinear characteristics of the oscillation. In this section, we describe how to identify the nonlinearity and distinguish its intensity.

#### A. Nonlinear control-loop performance for the VSC

It was shown in [4] and [21] that the amplitude limit in the control system is related to the oscillation caused by the VSC control. In this section, this type of oscillation caused by the control limit is referred to as "control oscillation". Figure 5 illustrates the block diagram of the VSC control, which conforms to the principle of double-PI control described in Section II, where the control limit can be modeled by saturation elements attached to the PI controllers and PWM module.

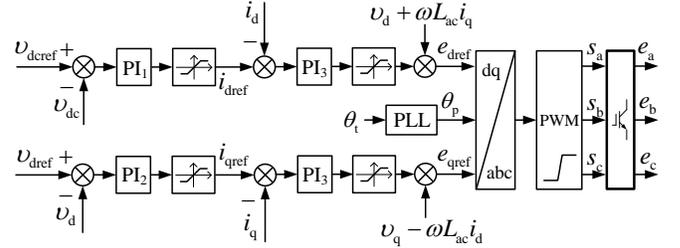

Fig. 5. Block diagram of the VSC control system, where $\theta_t$ is the phase angle of the terminal voltage at PCC and $\theta_p$ is the output of the phase-locked loop, which traces $\theta_t$ and can be used to specify the angle for the dq transformation.

In this paper, considering that the procedure delay of generating the internal potentials $e_a$, $e_b$, and $e_c$ from the control outputs $e_{dref}$ and $e_{qref}$ relies mainly on the dynamic of phase-locked loop (PLL), the controlling subsystem composed of the PLL, dq transformation (DQT), and PWM modules, as shown in Fig. 5, could be simplified to a proportion-inertia loop $K_{PWM}/(\tau_1 s + 1)$, where $K_{PWM}$ represents the equivalent gain for the subsystem and $\tau_1$ denotes the time constant. In addition, for the VSC control, the current close-loop tracking between the feedbacks ($i_d$, $i_q$) and references ($i_{dref}$, $i_{qref}$) can be modeled by the inertia loop with a small time constant $G_{ci}(s) = 1/(\tau_2 s + 1)$, because the inner current loop adjusts much faster than the outer voltage loop. Thus, Fig. 6 shows a simplified closed-loop block diagram for the d-axis control of the VSC, where

$$G_1(s) = K_{Pvd} + K_{Ivd}/s$$
$$G_2(s) = (1 - G_{ci}(s))(K_{Pi} + K_{Ii}/s)$$
$$G_3(s) = \dfrac{3v_d/2v_{dc}}{3L_{ac} C_{dc} s^2/2 + 3R_{ac} C_{dc} s/2 + (3v_d/2v_{dc})^2} \quad (28)$$

$$N_i(A_i) = \dfrac{2}{\pi} \left[ \arcsin\left(\dfrac{\delta}{A_i}\right) + \dfrac{\delta}{A_i}\sqrt{1 - \left(\dfrac{\delta}{A_i}\right)^2} \right]$$

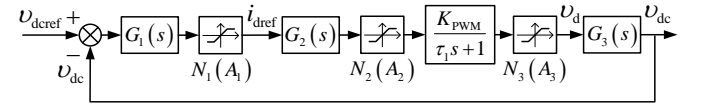

Fig. 6. Simplified closed-loop block diagram for the d-axis control of the VSC, where $N_i(A_i)$ ($i = 1,2,3$) are the describing functions of the saturation elements.

According to the principle of harmonic balance for nonlinear oscillation analysis, the system in Fig. 6 produces self-sustained oscillation if (29) has a solution.

$$1 + (K_{PWM}/(j\tau_1 \omega + 1)) \cdot \prod_{i=1,2,3} G_i(j\omega) \cdot \prod_{i=1,2,3} N_i(A_i) = 0 \quad (29)$$

Further, according to (29), we have

$$\begin{cases} 20\lg H_l(\omega) = -\sum_{i=1}^{3} 20\lg N_i(A_i) \\ \theta_l(\omega) = -\pi \end{cases} \quad (30)$$

where

$$H_l(\omega) = \left| K_{\text{PWM}} / (j\tau_1\omega + 1) \right| \cdot \prod_{i=1,2,3} |G_i(j\omega)|$$

$$\theta_l(\omega) = \sum_{i=1}^{3} \arg G_i(j\omega) + \arg\left(K_{\text{PWM}} / (j\tau_1\omega + 1)\right). \quad (31)$$

If the amplitude- and phase-frequency characteristics of (30) satisfy the inequalities [24]

$$\left. \frac{d(\lg H_l(\omega))}{d\omega} \right|_{\omega_0} \bigg/ \left. \frac{d\theta_l(\omega)}{d\omega} \right|_{\omega_0} > 0, \ \left. \frac{d(\lg N_i(a))}{da} \right|_{A_i} < 0, \quad (32)$$

the self-sustained solution $(A_i, \omega_0)$ obtained by (29) is stable.

Table I lists the parameters for the system shown in Fig. 6, and Fig. 7 shows the corresponding curves of the amplitude- and phase-frequency characteristics. As shown in Fig. 7, the linear part of the system exhibits low-pass filtering characteristics, and the frequency of the stable self-sustained solution is 35.7 Hz. Thus, if there is an influence of sustainable voltage disturbance on the control loop in Fig. 6, nonlinear oscillation occurs because of the control limit.

TABLE I
PARAMETERS OF THE MAIN CIRCUIT AND CONTROL FOR THE VSC

| Parameters | Values |
|---|---|
| DC-bus voltage control loop ($K_{Pvd}$, $K_{Ivd}$) | 2.5 pu, 1000 pu |
| Current control loop ($K_{Pi}$, $K_{Ii}$) | 50 pu, 6250 pu |
| Resistance and inductance for VSC ($R_{ac}$, $L_{ac}$, $C_{dc}$) | 1.224 Ω, 39.11 mH, 300 uF |
| Equivalent gain ($K_{\text{PWM}}$) | 0.353 pu |
| Equivalent time constant ($\tau_1$, $\tau_2$) | 0.00005 pu, 0.0003 pu |

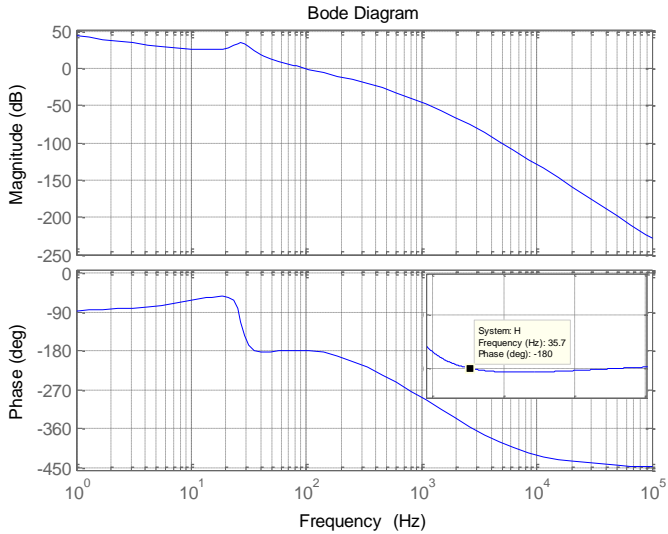

Fig. 7. Bode plot for $G_1(s) \cdot G_2(s) \cdot G_3(s)$.

### B. Index μ for Nonlinearity Detection

Reviewing Fig. 6, the self-sustained oscillation could be observed through the currents of the VSC because the currents are the variables in the forward channel of the VSC control. In view of timing signal analysis, the oscillatory current of the VSC conforms to the characteristic of a stationary random process $X(n) = A\cos(\Omega n + \Phi)$, where $A$, $\Omega$, and $\Phi$ are independent random variables, and $\Phi$ conforms to the uniform distribution ranging from 0 to $2\pi$; hence, the mathematical expectation of $X(n)$ is zero.

Furthermore, according to the principle of higher-order statistics, the third-order cumulant for the stationary random process $X(n)$, which is understood as the skewness coefficient, can be expressed as

$$c_{3(X)}(\tau_1, \tau_2) = E[X(n), X(n+\tau_1), X(n+\tau_2)], \quad (33)$$

and if $X(n)$, $X(n+\tau_1)$, and $X(n+\tau_2)$ are independent,

$$c_{3(X)}(\tau_1, \tau_2) = \begin{cases} \gamma_k, & \tau_1 = \tau_2 \\ 0 & \text{otherwise}. \end{cases} \quad (34)$$

In (34), it is shown that the cumulant $c_3(\tau_1, \tau_2)$ is an impulse function, and the Fourier transform of the impulse function is a constant; therefore, the spectrum of the cumulant $c_3$ is flat. In addition, the cumulant has a linear superposition property; that is, if the random processes $X(n)$ and $Y(n)$ are independent, then the cumulant

$$c_{3(X+Y)}(\tau_1, \tau_2) = c_{3(X)}(\tau_1, \tau_2) + c_{3(Y)}(\tau_1, \tau_2). \quad (35)$$

Thus, if a signal $X(n)$ comprising the sinusoidal fundamental and harmonic components satisfies the principle of linear superposition, the cumulant of the signal is equal to the sum of the cumulants for each component, and according to (34), the spectrum of the third-order cumulant for $X(n)$ is flat. However, if there is coupling between the harmonic components of $X(n)$ such that the signal $X(n)$ is nonlinear, i.e., it does not satisfy the superposition principle, the resulting spectrum of the cumulant $c_{3(X)}$ will not be flat. Hence, the nonlinearity hidden in signal $X(n)$ can be detected using cumulant calculations.

Moreover, in practice, the spectrum of $c_{3(X)}$ is usually defined as a bispectrum $B_X(\omega_1, \omega_2)$, and the bispectrum can be normalized to an absolute scale in the range 0 through 1, which is called bicoherence coefficient. The spectra of the bicoherence coefficient and its magnitude are defined as follows:

$$K_{2X}(f_1, f_2) = \frac{B_X(f_1, f_2)}{\sqrt{P_X(f_1) P_X(f_2) P_X(f_1+f_2)}} \quad (36)$$

$$bic \triangleq |K_{2X}(f_1, f_2)|$$

where $f_1$, $f_2$, and $(f_1+f_2)$ are the frequencies of the Fourier transformation; and $P_X(\cdot)$ represents the power spectrum. According to (36), the bicoherence spectrum could reflect a coupling phenomenon between the components at the frequencies $f_1$ and $f_2$. The square of *bic* could be proven to represent the fraction of the power generated by the nonlinear coupling between the components at $f_1$ and $f_2$ to the total power of the component at $(f_1+f_2)$. Thus, according to the property of cumulant calculation, the criterion of nonlinearity detection for a random process $X(n)$ can be described as follows: If the value of *bic* for $X(n)$ is constant, the process $X(n)$ is linear; otherwise, the situation is the opposite. Therefore, a nonlinearity index $\mu$ for verifying the flatness of *bic* could be defined as

$$\mu \triangleq \left| \hat{K}_{2X,\max}^2 - \left( \overline{\hat{K}_{2X}^2} + 2\sigma_{\hat{K}_{2X}^2} \right) \right| \quad (37)$$

where $\hat{K}_{2X,max}^2$ represents the estimation of the maximum squared bicoherence; and $\sigma_{\hat{K}_{2X}^2}$ is the variance of $\hat{K}_{2X}^2$. In (37), if $\mu > 0$, the signal generating process is nonlinear.

However, the above criterion is theoretically effective, and the actual results obtained by measurements and calculations may violate this criterion. The reasons include: 1) The measurement data are a finite-length segment for an actual signal, and the results of *bic* derived from the finite-length data are also affected by the FFT parameter; 2) for the power systems, the actual voltage and currents contain background harmonics whose content rates are allowed by the power-quality standards; however, the signals with harmonics would affect the conclusion of the nonlinearity detection. Therefore, a threshold that considers an actual signal with background harmonics must be discussed.

In this study, the nonlinearity detection of VSC control is based on the measurement of output currents with background harmonics. According to the power-quality standard [25], Table II reports the harmonic limitations for 35-kV networks, and the corresponding result of index $\mu$ is 0.0180 obtained by simulation; compared with the result from simulation, the threshold value has sufficient margin.

TABLE II
HARMONIC LIMITATION AND THE THRESHOLD FOR NONLINEARITY DETECTION

| Harmonic order | $THD_U$ limitation | Result of $\mu$ | Threshold of $\mu$ |
|---|---|---|---|
| 2~25 | 3% for voltage | 0.0180 | 0.10 |

## IV. PROPOSED SCHEME FOR OSCILLATION SOURCE LOCATION

To locate the oscillation sources, the previous sections presented the energy structure of the VSC and a nonlinearity index for the VSC control. Accordingly, this section presents a comprehensive scheme for OSL. Fig. 8 shows a concise flowchart of the proposed scheme, whose details are as follows:

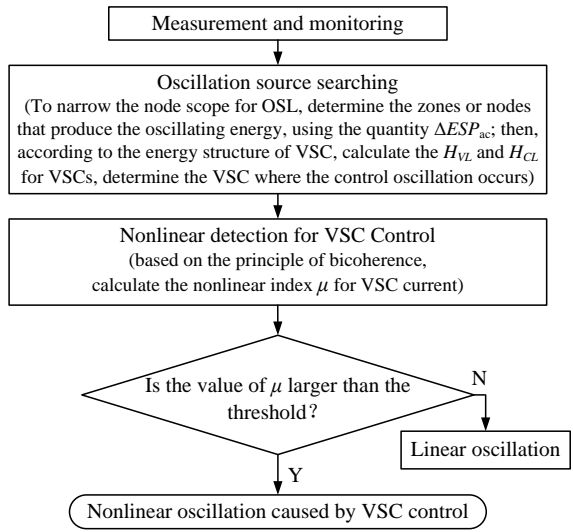

Fig. 8. Flowchart of source location for VSC oscillation.

*Step 1: Measurement and oscillation monitoring.*

Acquire the sampling sequences of the instantaneous voltages and currents for nodes in the networks, and calculate the characteristic quantity, $\Delta ESP_{ac}$, defined by (16); if $\Delta ESP_{ac}$ increases continuously, it is demonstrated that oscillatory phenomena occur in the system; then, go to Step 2.

*Step 2: Oscillation source searching.*

To narrow the node scope for locating the possible zones or nodes which produce the oscillation energy, filter the nodes that are definitely not oscillation sources via listing the values of $\Delta ESP_{ac}$ in descending order and determining whether the characteristic quantity $\Delta ESP_{ac}$ is no more than zero. Then, for the VSC nodes on the above list, analyze the Hamiltonian storage functions $H_{VL}$ and $H_{CL}$, and if the values of $H_{VL}$ and $H_{CL}$ increase continuously, determine the VSC node as an oscillation source. Evidently, the VSC nodes that generate the oscillation energy caused by the control limit are included in the list.

*Step 3: Determine whether the nonlinear oscillation occurs for the VSC*

According to the result of the oscillation source searching obtained in Step 2, calculate the value of nonlinearity index $\mu$ by measuring the output current of the VSC. If the $\mu$ value for the VSC is higher than the threshold, the oscillation associated with the VSC is nonlinear.

## V. CASE STUDY

Comparing with the existing literature, the features of the proposed methodology for locating the VSC oscillation source are: (i) A characteristic quantity $\Delta ESP_{ac}$ associated with TEF is used to monitor the oscillation and narrow the node scope for the OSL; (ii) The Hamiltonian energy of the VSC control is applied to search the VSCs which produce energy in the lower level networks, cooperating in the method of TEF; (iii) The method of nonlinearity detection is involved in determining the type of the oscillation sources, i.e., linear or nonlinear oscillation. The following case study focuses on these points.

Fig. 9 illustrates the topology of the case-study system derived from the IEEE 9-bus system, where the parameters of the network composed of buses 1 to 9 are consistent with the IEEE 9-bus system. In contrast to the IEEE benchmark system, the system shown in Fig. 9 locates an SVC at medium-voltage bus 13 for reactive power adjustment, and grid-connected wind farms and their corresponding static var generators (SVGs) are located on buses 14 and 15. The SVGs adopt double-loop PI control, as described in Section II-B, and the control parameters are listed in Table III.

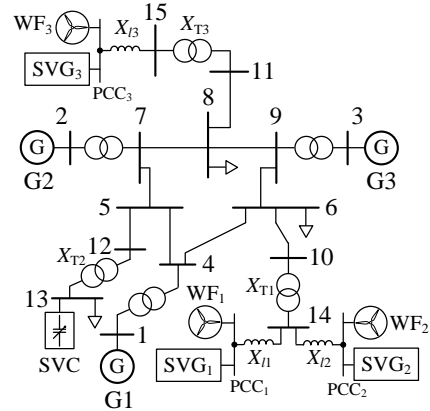

Fig. 9. The topology of the case study system.

TABLE III
PARAMETERS OF NETWORK AND VSC CONTROL FOR CASE-STUDY SYSTEM

| Parameters | Values |
| --- | --- |
| SVG voltage control loop ($K_{Pvd}$, $K_{Ivd}$, $K_{Pvq}$, $K_{Ivq}$) | 2.5 pu, 1000 pu, 2 pu, 20 pu |
| Reference of terminal voltage control ($V_{ref1}$, $V_{ref2}$, $V_{ref3}$) | 1.005 pu, 1.005 pu, 1.005 pu ($t < 2$ s) 1.005 pu, 1.000 pu, 1.005 pu ($t \geq 2$ s) |
| Current control loop ($K_{Pi}$, $K_{Ii}$) | 40 pu, 6250 pu |
| Connection impedance ($X_{l1}$, $X_{l2}$, $X_{l3}$) | 0.0051 pu, 0.0038 pu, 0.0256 pu |
| Line resistance ($R_{6-10}$, $R_{8-11}$) | 0.0017 pu, 0.0054 pu |
| Line impedance ($X_{6-10}$, $X_{8-11}$) | 0.0092 pu, 0.0178 pu |
| Transformer impedance ($X_{T1}$, $X_{T2}$, $X_{T3}$) | 0.0586 pu, 0.0586 pu, 0.0576 pu |

The case-study system was implemented using PSCAD/EMTDC simulation. Fig. 10 shows the waveforms of the voltages and currents. In Fig. 10 (a), when $t < 2$ s, the voltage amplitudes $V_{14}$, $E_{SVG1}$, and $E_{SVG2}$ are stable; by contrast, when $t \geq 2.45$ s and $V_{ref1} \neq V_{ref2}$, the voltage amplitude $V_{14}$ deviates from the previous equilibrium point, which fluctuates and ranges from 0.97 to 1.04 pu. Meanwhile, according to Fig. 10 (b), the current amplitude of $i_{14}$ increases significantly when $t \geq 2.45$ s, and the total harmonic distortion (THD) of $i_{14}$ is up to 249.11%. The major harmonic frequencies of instantaneous current $i_{14}$ are 17.5 and 82.5 Hz (50 ± 32.5 Hz); therefore, it is demonstrated that there is a sub-synchronous current injection flowing from VSCs to networks.

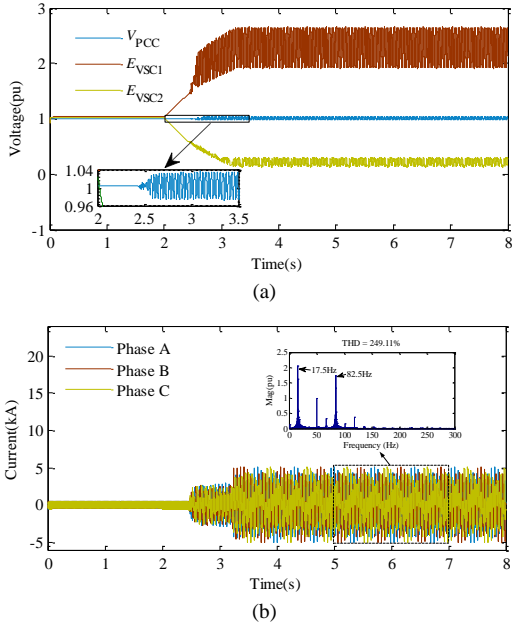

Fig. 10. Waveforms of voltages and currents. (a) Voltage amplitudes. (b) Three-phase currents $i_{14}$ and the corresponding spectrum.

Further, according to (13) and (16), Fig. 11 illustrates the curves of the energy function $V$ and characteristic quantity $\Delta ESP_{ac}$. As shown in Fig. 11, when $t \geq 2.45$ s and $V_{ref1} \neq V_{ref2}$, the curves of $V$ for SVG$_1$ and SVG$_2$ are both increasing, which demonstrates that the oscillation energy is produced by SVG$_1$ and SVG$_2$; meanwhile, the curve of $\Delta ESP_{ac}$ also increases when $t \geq 2.45$ s ($\Delta ESP_{ac} > 0$), and its growth trend is similar to that of $V$. Thus, the characteristic $\Delta ESP_{ac}$ can reflect the properties of oscillation. Further, Fig. 11 shows the curve of the other characteristic $W_{TEF} = \int i_x dv_y - i_y dv_x$ derived from [20], however, the changing trend of $W_{TEF}$ is opposite to that of $V$. Hence, the result in Fig. 11 verifies the effectiveness of the proposed characteristic quantity $\Delta ESP_{ac}$, which can be applied to determine the zone including the possible SVGs that produce the oscillation energy.

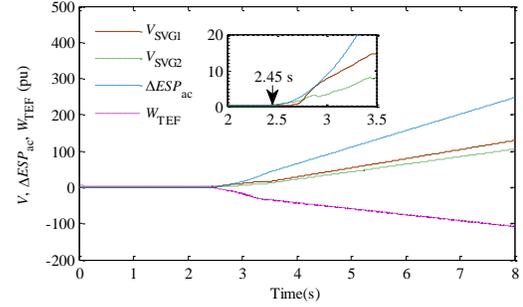

Fig. 11. Curves of the energy function $V$ and the characteristic quantity $\Delta ESP_{ac}$.

Fig. 12(a) shows the waveforms of the Hamiltonian energy. In Fig. 12(a), the energy curves indicate that SVG$_1$ and SVG$_2$ produce oscillation energy when $t \geq 2.45$ s because of $H_{VL} + H_{CL} > 0$; by contrast, the Hamiltonian energy of SVG$_3$ is almost equal to 0. It is shown that, cooperating in the identification of the characteristic quantity $\Delta ESP_{ac}$, the analysis of the Hamiltonian energy could further determine the SVGs where the control oscillation occurs.

Nonlinearity detection was performed according to the proposed scheme, as shown in Fig. 8. Reviewing (5), the Hamiltonian energy of the VSC is related to inductance $L_{ac}$ and capacitance $C_{dc}$. However, inductance and capacitance elements are very common in networks; therefore, it is necessary to distinguish between the LC resonance and VSC control oscillation in the OSL. Thus, a linear oscillation was triggered to verify the effectiveness of the proposed nonlinearity index. Fig. 12(b) shows the current waveform for the SVC at bus 13. The combination of the SVC capacitive reactance and network inductance matches the sub-synchronous frequency of $i_{SVG1}$ and $i_{SVG2}$ therefore, the SVC stimulates a linear LC resonance. Fig. 12(b) shows that the oscillation diverges and the SVC works as a harmonic amplifier.

Table IV reports the results of nonlinearity detection using the index $\mu$. In the table, the $\mu$ values for SVG$_1$ and SVG$_2$ are significantly higher than that for SVC, implying that both SVGs are the major contributors to the nonlinearity. The result of $\mu$ is essentially consistent with the trend of the Hamiltonian energy, as shown in Fig. 12(a), i.e., the converter with a large Hamiltonian energy has a large nonlinearity index $\mu$. By contrast, if we locate the nonlinear oscillation sources according to THD$_I$ (i.e., the power converter, which produces a self-sustained oscillation, is regarded as a harmonic source), the result will be opposite, because the values of THD$_I$ could not represent the nonlinear characteristic of oscillation. Thus, the proposed index $\mu$ is of benefit to filter the nodes that are definitely linear oscillation sources.

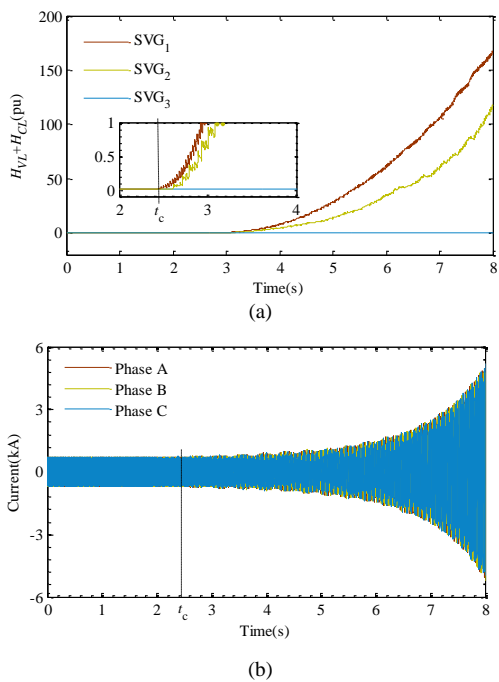

Fig. 12. Waveforms of Hamiltonian energy and currents. (a). Hamiltonian energy for SVG control. (b). current waveform for the SVC.

TABLE IV
VALUE COMPARISON OF $\mu$ AND $THD_I$ FOR HARMONIC SOURCE SEARCHING

| Index | $SVG_1$ | $SVG_2$ | SVC |
|---|---|---|---|
| $\mu$ (pu) | 0.3727 | 0.1598 | 0.0079 |
| $THD_I$ (%) | 77.2610 | 59.5503 | 385.3142 |

Therefore, the results of the case study confirm that the proposed scheme for OSL can locate the voltage-source converters that cause control oscillations and distinguish the contributions of different converters to the oscillation energy.

## VI. CONCLUSION

This study proposes a scheme for locating the source of the SSO caused by the control of the VSC based on the energy structure and nonlinearity detection. First, the energy structure of the VSC is proposed via PCH modeling. According to the energy structure, on one hand, the characteristic quantity $\Delta ESP_{ac}$ is defined to implement the oscillation monitoring and narrow down the node scope for OSL; and on the other hand, the Hamiltonian energy functions $H_{VL}$ and $H_{CL}$ are applied to determine the converter where the control oscillation occurs. To further determine the type of oscillation and identify the oscillation responsibility, a nonlinearity index and its threshold are discussed. The oscillatory power converter is located based on the proposed scheme for OSL. Finally, the results of a case study implemented using the PSCAD/EMTDC simulation validates the proposed scheme.